# Adaptive model selection in photonic reservoir computing by reinforcement learning

**Kazutaka Kanno**[1,*], **Makoto Naruse**[2], **and Atsushi Uchida**[1]

[1]Department of Information and Computer Sciences, Saitama University 255 Shimo-Okubo, Sakura-ku, Saitama City, Saitama, 338–8570 Japan

[2]Department of Information Physics and Computing, Graduate School of Information Science and Technology, University of Tokyo, 7-3-1 Hongo, Bunkyo-ku, Tokyo, 113-8654 Japan

*kkanno@mail.saitama-u.ac.jp

**ABSTRACT**

Photonic reservoir computing is an emergent technology toward beyond-Neumann computing. Although photonic reservoir computing provides superior performance in environments whose characteristics are coincident with the training datasets for the reservoir, the performance is significantly degraded if these characteristics deviate from the original knowledge used in the training phase. Here, we propose a scheme of adaptive model selection in photonic reservoir computing using reinforcement learning. In this scheme, a temporal waveform is generated by different dynamic source models that change over time. The system autonomously identifies the best source model for the task of time series prediction using photonic reservoir computing and reinforcement learning. We prepare two types of output weights for the source models, and the system adaptively selected the correct model using reinforcement learning, where the prediction errors are associated with rewards. We succeed in adaptive model selection when the source signal is temporally mixed, having originally been generated by two different dynamic system models, as well as when the signal is a mixture from the same model but with different parameter values. This study paves the way for autonomous behavior in photonic artificial intelligence and could lead to new applications in load forecasting and multi-objective control, where frequent environment changes are expected.

**Introduction**

Reservoir computing involves information processing based on recurrent neural networks[1]. This method is known to be suitable for temporal or sequential information processing, such as time series prediction[2] and speech recognition[3]. In reservoir computing, the input data to be processed are fed into a recurrent neural network, which is called a reservoir. The reservoir network produces a transient response when the input signal is injected. The reservoir computing processing result is the weighted linear sum of the node states in the reservoir. The main characteristic of reservoir computing is that the input weights and reservoir are fixed, being specified by the physical characteristics of the reservoir, while the output weights are trained. These characteristics significantly reduce the computational cost of learning compared with those of standard recurrent neural networks.

Nonlinear mapping of the input data into a high-dimensional space is required to achieve reservoir functionality for successful computation[4]. This functionality can be realized using other nonlinear dynamic systems instead of recurrent neural networks. Reservoir computing based on various types of nonlinear dynamic systems has been proposed[5–9]. Photonic implementation of reservoir computing is one example, where a semiconductor laser with a delayed feedback loop is used as a reservoir[10–13]. One of the advantages of photonic reservoir computing is that it enables the realization of fast information processing with low learning cost using established optoelectronic devices. It has been reported that speech recognition at a rate of 1.1 Gb/s can be achieved using photonic reservoir computing[12].

Reservoir computing can, however, only adapt to input signals that are used to train the output weights of the reservoir. In other words, reservoir computing does not work well if the incoming signals do not correspond with the training datasets. In reality, environmental conditions may change the characteristics of the observations, which could induce variations of the input that are different from the original knowledge used in the training phase. Additionally, it is assumed that the input signals could be generated by many different dynamic source models and the source model is dynamically switched in time or the signal is a mixture of different source models. It may be difficult to train the reservoir computing system to produce the correct outputs for *all* different models or arbitrary environmental conditions.

To solve this serious issue, we propose a scheme of reservoir computing combined with reinforcement learning in this study. In this scheme, training is conducted with respect to individual input signals generated by a designated model. Hence, multiple output weights of the reservoir are obtained, corresponding to the different types of source signals in the training phase. In the task execution phase, one of the output weights of the reservoir is selected such that the minimum prediction error for the given input signals is achieved by reinforcement learning. This adaptive model selection scheme is expected to be useful for applications such as load forecasting[14], multi-objective control[15], and signal recovery in communication[16] when environmental changes or diverse types of input signals are expected; hence, the preparation of multiple output weights of the reservoir prior to execution and dynamic model selection would be highly effective.

Decision making using reinforcement learning is a machine learning scheme concerned with the problem of training an action policy to maximize the total reward[17]. The multi-armed bandit (MAB) problem is a fundamental problem in reinforcement learning, whose goal is to maximize the total reward when agents select one of multiple slot machines with unknown hit probabilities in finite trials[17]. The idea of adaptive model selection stems from associating the slot machines in the MAB problem with the trained output weights of the reservoir. Therefore, the strategy used to solve the MAB problem[18–20] could be effective in adaptive model selection. Furthermore, several methods of photonic decision making have been demonstrated with operation in the gigahertz regime by utilizing chaotic laser time series[21]. Notably, both reservoir computing and dynamic model selection can be performed on a photonic platform for ultrafast operation[22,23].

In this study, we numerically demonstrate adaptive model selection using decision making based on chaotic laser outputs in photonic reservoir computing with reinforcement learning. We consider a situation in which the input signal is generated by one of two dynamic models, specifically, the Lorenz model[24] or Rössler model[25], and the input signal is switched in time between the two models to mimic environmental changes. We train the reservoir using the time series generated by either one of the two models and prepare two types of output weights for the reservoir corresponding to the two models. We perform time series prediction of the input signal using reservoir computing. Generally, if the output weights of a reservoir do not correspond to the characteristics of the actual input signals, for instance due to environmental changes, a larger prediction error is obtained. In reinforcement learning, action policies are trained based on rewards, and the prediction errors in reservoir computing are regarded as rewards in this study. The proposed scheme autonomously changes the output weights of the reservoir according to the given input signals to reduce the prediction error. We numerically demonstrate correct adaptive model selection for different configurations of the dynamic models.

**Adaptive model selection based on decision making in photonic reservoir computing**

We propose a scheme for adaptive model selection based on decision making in photonic reservoir computing. Figure 1 schematically illustrates the architecture of the proposed approach. The scheme comprises three parts: photonic reservoir computing, reinforcement learning, and generation of chaotic laser outputs. In this study, we numerically implement photonic reservoir computing and reinforcement learning. We use experimentally generated chaotic temporal waveforms of the laser outputs for reinforcement learning in the numerical simulations. The photonic reservoir computing system consists of a semiconductor laser with optical feedback. (See the *Methods* section for details.) In this scheme, chaotic time series prediction is numerically performed using photonic reservoir computing, where a predicted signal is generated using two dynamical models: the Lorenz[24] and Rössler models[25]. Considering the situation in which the source of the input signal changes over time, mimicking environmental changes, single-point prediction is performed using photonic reservoir computing. Two types of reservoir output weights are prepared, which are trained by chaotic time series generated separately using the Lorenz and Rössler models. Two predicted time series are generated based on the two output weights. In the adaptive model selection, the prediction errors for the two output weights are utilized with the objective of determining which model should be used for time series prediction.

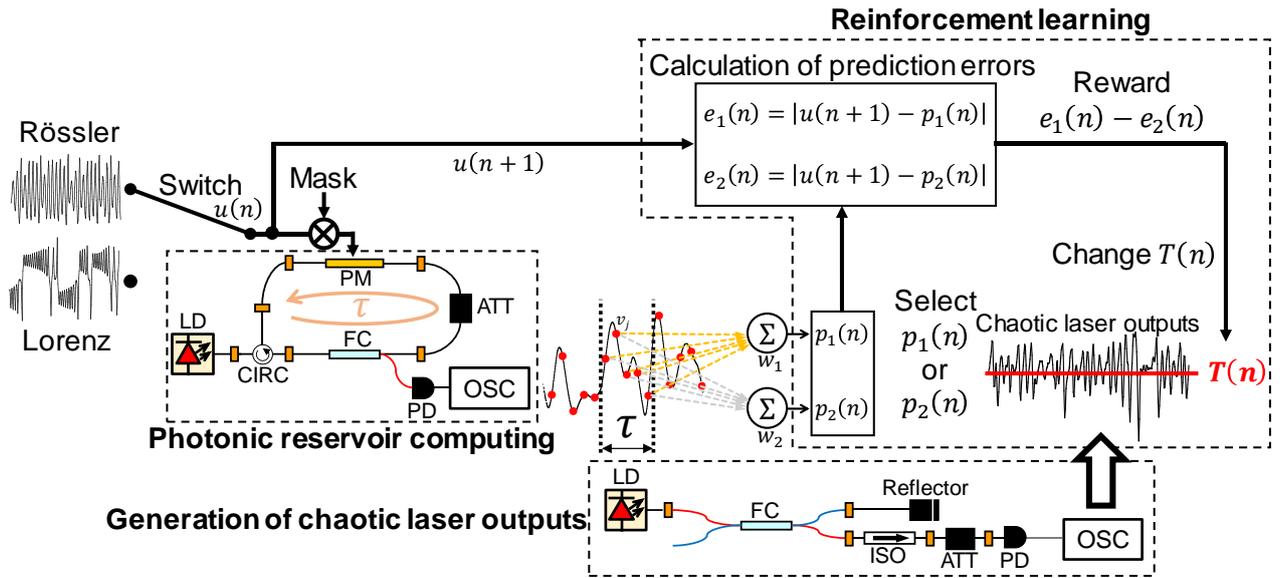

**Figure 1.** Schematic diagram of adaptive model selection using reservoir computing and reinforcement learning. The system comprises three parts: photonic reservoir computing, reinforcement learning, and chaotic laser system. LD is laser diode, PM is phase modulator, CIRC is optical circulator, ATT is optical attenuator, FC is optical fiber coupler, PD is photodetector, ISO is optical isolator, and OSC is digital oscilloscope.

The input chaotic time series is denoted by $u(n)$ (Fig. 1). The task of the reservoir is to conduct a single-point prediction of $u(n)$; that is, the reservoir computing predicts $u(n+1)$ when $u(n)$ is injected into the reservoir. Two types of output weights are trained separately using the time series from the two models and are represented as $w_1$ and $w_2$. The reservoir produces two predicted outputs, $p_1(n)$ and $p_2(n)$, using output weights $w_1$ and $w_2$, respectively.

Adaptive model selection is performed by determining whether the prediction error $e_1(n)$ or $e_2(n)$ is smaller.

The method of decision making based on chaotic laser output is employed to select one of the two output weights[21,23]. In this method, a chaotic laser time series generated by a semiconductor laser with optical feedback is used[26], and the sampled data of the chaotic output are compared with a threshold value $T(n)$. If the chaotic laser output is larger (smaller) than $T(n)$, $p_1(n)$ ($p_2(n)$) is selected. Then, we go to the next time step ($n \rightarrow n + 1$) and use the next input datum $u(n + 1)$. From $u(n + 1)$, $p_1(n)$, and $p_2(n)$, $e_1(n)$ and $e_2(n)$ can be obtained using $e_{1,2}(n) = |u(n + 1) - p_{1,2}(n)|$. The smaller prediction error is determined by comparing $e_1(n)$ and $e_2(n)$, and $T(n)$ is changed accordingly. If $e_1(n)$ is smaller (larger) than $e_2(n)$, then $T(n)$ is decreased (increased). The change in $T(n)$ increases the probability of selecting the predicted output with the smaller error. By repeating the change in $T(n)$ based on the comparison of $e_1(n)$ and $e_2(n)$, $T(n)$ becomes much smaller or larger than the probability distribution of the chaotic laser output, and only one of $p_1(n)$ or $p_2(n)$ is selected. Then, the correct values of $w_1$ and $w_2$ for adaptive model selection are determined.

$T(n)$ is changed via the threshold adjuster $TA(n)$ and is defined as follows:

$$T(n) = \begin{cases} kN_T & (TA(n-1) > N) \\ k\lfloor TA(n) \rfloor & (-kN_T \leq T(n) \leq kN_T. \\ -kN_T & (TA(n-1) < -N_T) \end{cases} \quad (1)$$

$\lfloor TA(n) \rfloor$ is the nearest integer to $TA(n)$ rounded to 0. In this study, $\lfloor TA(n) \rfloor$ was assumed to take the values $-N_T, \ldots -1, 0, 1 \ldots, N_T$, where $N_T$ is a natural number. Hence, the number of thresholds is $2N_T + 1$. The threshold number and $k$ in Eq. (1) determine the range of $T(n)$. The range of $T(n)$ is limited from $-kN_T$ to $kN_T$ by setting $T(n) = kN_T$ when $TA(n) > N_T$ and $T(n) = -kN_T$ when $TA(n) < -N_T$. $TA(n)$ is changed based on the relationship between the magnitudes of $e_1(n)$ and $e_2(n)$ as follows:

$$TA(n+1) = \begin{cases} \alpha TA(n) - 1 & (e_1(n) \leq e_2(n)) \\ \alpha TA(n) + 1 & (e_1(n) > e_2(n)) \end{cases}, \quad (2)$$

where $\alpha$ is referred to as the forgetting (memory) parameter[27,28]. A large value of $\alpha$ means that the dynamics of $TA(n)$ holds memory of the initial value of $TA(n)$. In this scheme, the sum of the hit probabilities of the two slot machines (models) is supposed to be fixed at 1, because one of the two models is always selected. Therefore, the threshold shift is fixed at 1. A temporal waveform of chaotic laser outputs used for decision making was experimentally obtained from a semiconductor laser with optical feedback[21]. The semiconductor laser was subjected to delayed optical feedback by using an external fiber reflector, inducing chaotic temporal waveforms in the intensity of the laser output[26]. The chaotic output was detected using a photodetector and sampled by a high-speed digital oscilloscope. The sampling interval of the digital oscilloscope was 10 ps, and the chaotic laser output was sampled at this interval. In this study, decision making is performed at a sampling interval of 50 ps, because it has been reported that this sampling interval yields the best performance due to the existence of a negative correlation[21].

The vertical resolution of the digital oscilloscope was 8 bits, and the sampled data had 8-bit resolution. In the decision making method, the chaotic data sampled by the oscilloscope are compared to $T(n)$. We thus limited the range of $T(n)$ to $-128 \leq T(n) \leq 128$. To determine the shift of $T(n)$, $N_T = 8$ and $k = 16$ were used in this study. The number of threshold levels was $2N_T + 1 = 17$.

## Results and Discussion

**Adaptive model selection between Rössler and Lorenz models**

We numerically demonstrate adaptive model selection based on decision making in chaotic time series prediction. To generate a prediction target, we use two models, the Rössler and Lorenz models, which are well-known models that can produce chaotic behaviors (see the *Methods* section for details). A time series is generated using one of the two models, and the models are switched over time. Figure 2 shows the input signals produced by the two models. The first 500 points of the time series are generated by the Lorenz model, which is then switched to the Rössler model for the next 500 points. After that, the model is periodically switched every 500 points.

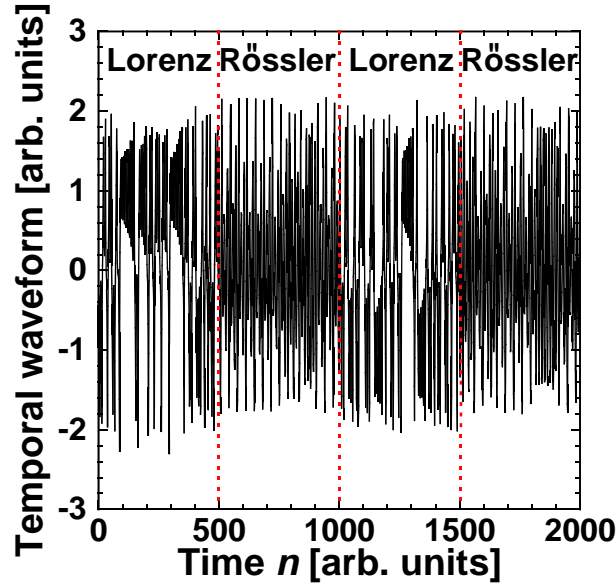

**Figure 2.** Temporal waveform generated using the Lorenz and Rössler models. The first 500 points of the waveform are produced using the Lorenz model, and the model is switched every 500 points.

The values of the output weights $w_1$ and $w_2$ of the reservoir for the Rössler and Lorenz models are presented in Figs. 3(a) and 3(d), respectively. The $i$-th element of the weight corresponds to the $i$-th virtual node in photonic reservoir computing. The time series generated using the Rössler and Lorenz models are employed to calculate $w_1$ and $w_2$, respectively, in the training procedure. The number of points used for training is 5,000, and the two weights are different from each other.

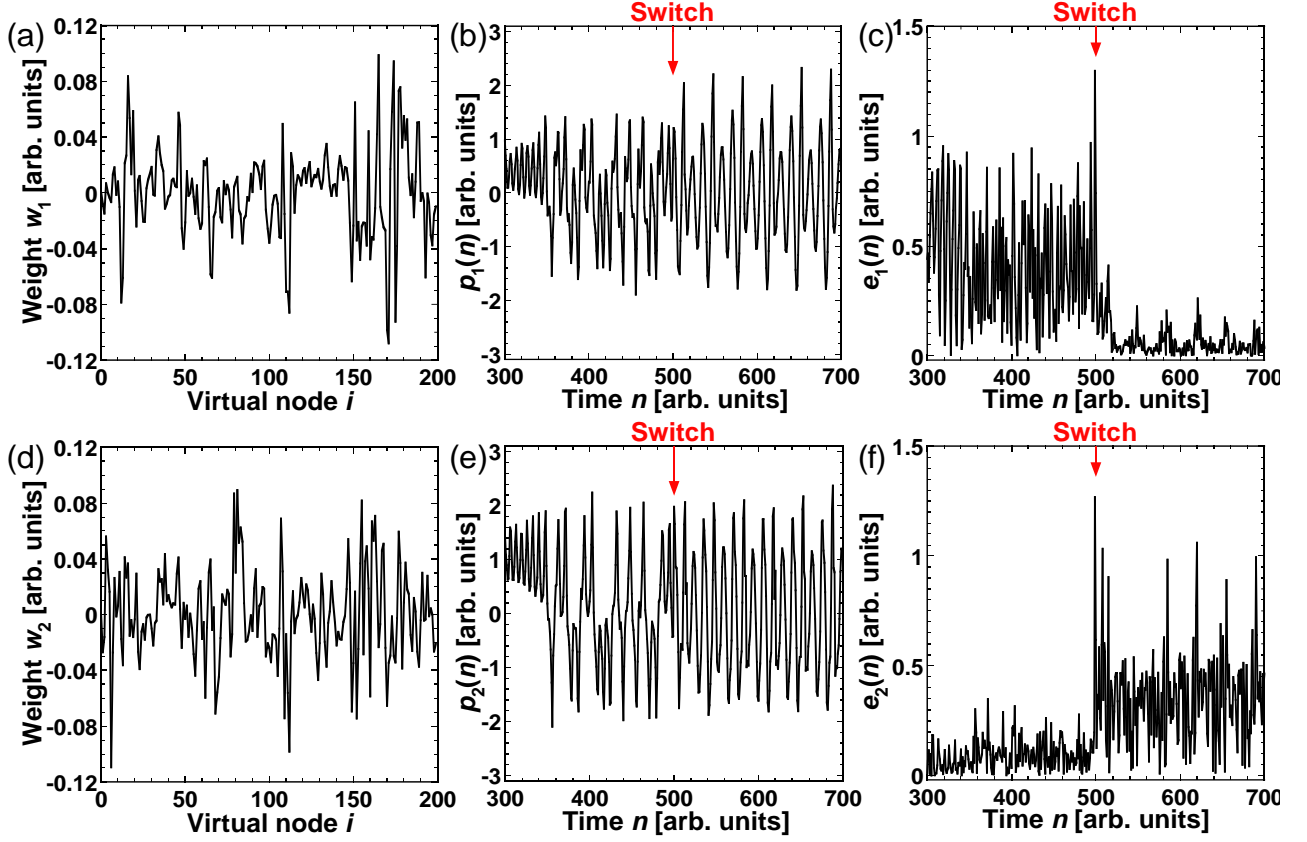

**Figure 3.** Prediction results obtained using reservoir computing. (a), (d) Trained output weights $w_1$ and $w_2$ of the reservoir for the Rössler and Lorenz models, respectively. (b), (e) Predicted time series obtained from $w_1$ and $w_2$, respectively. (c), (f) Prediction errors calculated from the differences between the input and predicted time series.

Figures 3(b) and 3(e) show the predicted time series of $p_1(n)$ and $p_2(n)$, which were generated using $w_1$ and $w_2$, respectively. Figures 3(c) and 3(f) depict the prediction errors $e_1(n)$ and $e_2(n)$, respectively. These figures are enlarged in $300 \leq n \leq 700$, which includes the switching of the time series from the Lorenz model to the Rössler model at $n = 500$. $e_2(n) < e_1(n)$ when $300 < n < 500$, where the prediction target is the Lorenz model. Meanwhile, $e_1(n) < e_2(n)$ when $500 < n < 700$, where the prediction target is the Rössler model. These results indicate that the error of a predicted waveform generated using the output weight corresponding to the prediction target is smaller.

An example of adaptive model selection is provided in Fig. 4, where the prediction target is the Lorenz model. Figure 4(a) shows the difference between the two errors, $\Delta e(n) = e_1(n) - e_2(n)$. The relationship between the magnitudes of the errors can be determined from $\Delta e(n)$. A positive value of $\Delta e(n)$ indicates that $e_2(n) < e_1(n)$ in Fig. 4(a). The temporal evolution of $T(n)$ is shown in Fig. 4(b). $T(n)$ for decision making varies based on $\Delta e(n)$ and increases to 128 after fluctuating around 0 at a small time step. The predicted output is selected by comparing $T(n)$ with the chaotic laser output, and the selection result is shown in Fig. 4(c). When $T(n)$ fluctuates around 0 in Fig. 4(b), either $p_1(n)$ or $p_2(n)$ may be selected. After $T(n)$ reaches 128, only $p_2(n)$ can be selected. Thus, the predicted output corresponding to the target (the Lorenz model) is selected successfully.

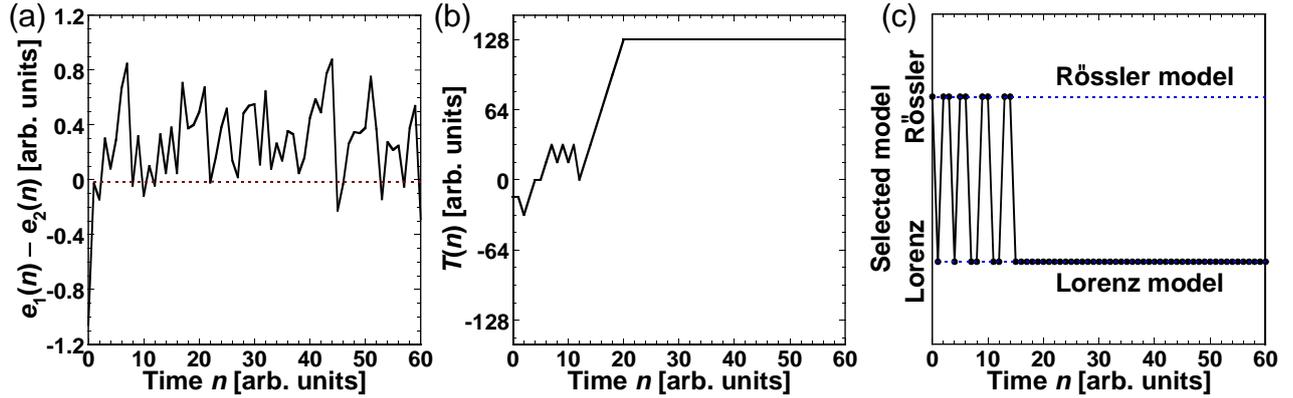

**Figure 4.** (a) Time series of the differences between two prediction errors $\Delta e(n) = e_1(n) - e_2(n)$. (b) Temporal dynamics of $T(n)$ for decision making. (c) Models selected by decision making in each step. The Lorenz model is selected for $n > 14$.

To investigate the adaptation of model selection to sudden environmental changes, we demonstrate time series prediction with model switching. The target time series shown in Fig. 2 consists of 2,000 points, so model selection is repeated 2,000 times. The prediction of the time series with 2,000 points is repeated 100 times. In each trial, different time series are generated by the two models and used as the prediction targets. We calculate the *correct model selection rate*, denoted by $\text{CMSR}(n)$, which is defined as the ratio of the number of selections of the predicted output corresponding to the target model at time $n$ among 100 trials. If $\text{CMSR}(n) = 1$, then the model used for time series prediction at time $n$ perfectly agrees with the original input signal source model.

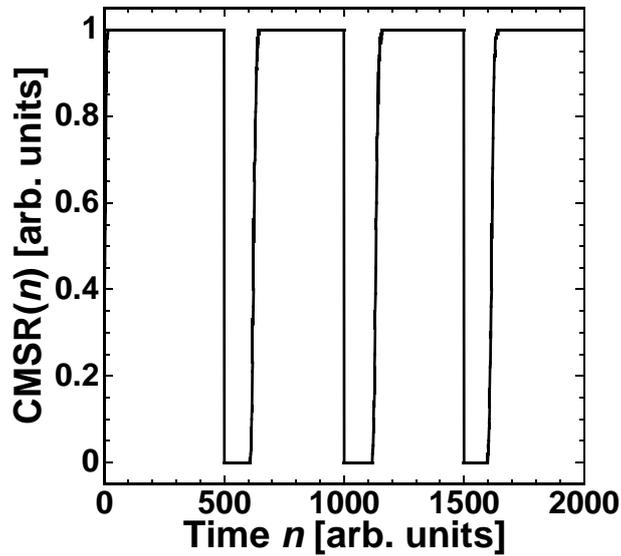

**Figure 5.** Correct model selection rate (CMSR) in adaptive model selection based on decision making in time series prediction. The models are switched between the Lorenz and Rössler models every 500 steps, as shown in Fig. 2.

Figure 5 shows the temporal evolution of $\text{CMSR}(n)$, which increases quickly to 1 after the prediction begins. When the target model is switched at $n = 500$, 1,000, and 1,500, CMSR($n$) decreases to 0. After the switch, CMSR($n$) increases to 1 again. Therefore, the correct model is selected adaptively under model switching (i.e., environmental changes). In addition, we note that the switching of the model selection may randomly occur in general situations. The cases of the model selection at different switching times are present in the Supplementary Information.

**Adaptive model selection with mixed time series from Rössler and Lorenz models**

Adaptive model selection in the Rössler and Lorenz models is a simple case because the difference between $e_1(n)$ and $e_2(n)$ is large, as shown in Fig. 4. In this subsection, a more difficult case of adaptive model selection is described, in which a mixed time series is used for the prediction target, as shown in Fig. 6(a). A mixed time series is generated by the Rössler and Lorenz models, where the two kinds of time series are mixed with different ratios. The two mixed time series are given by $x_1 = ax_L + (1-a)x_R$ and $x_2 = (1-a)x_L + ax_R$, where $x_R$ and $x_L$ represent the time series generated by the Rössler and Lorenz models, respectively, and the coefficient $a$ is the ratio of the Lorenz model in the mixed time series. The mixed time series is shown in Fig. 6(b), which is obtained with $a$ fixed to 0.8 and $x_1$ and $x_2$ used alternately every 500 points. The time series generated by the Rössler and Lorenz models are used to train $w_1$ and $w_2$, respectively. The aim of this model selection using the mixed time series is to select the time series corresponding to the model with the larger $a$, that is, $x_1$ for the Lorenz model and $x_2$ for the Rössler model at $a = 0.8$.

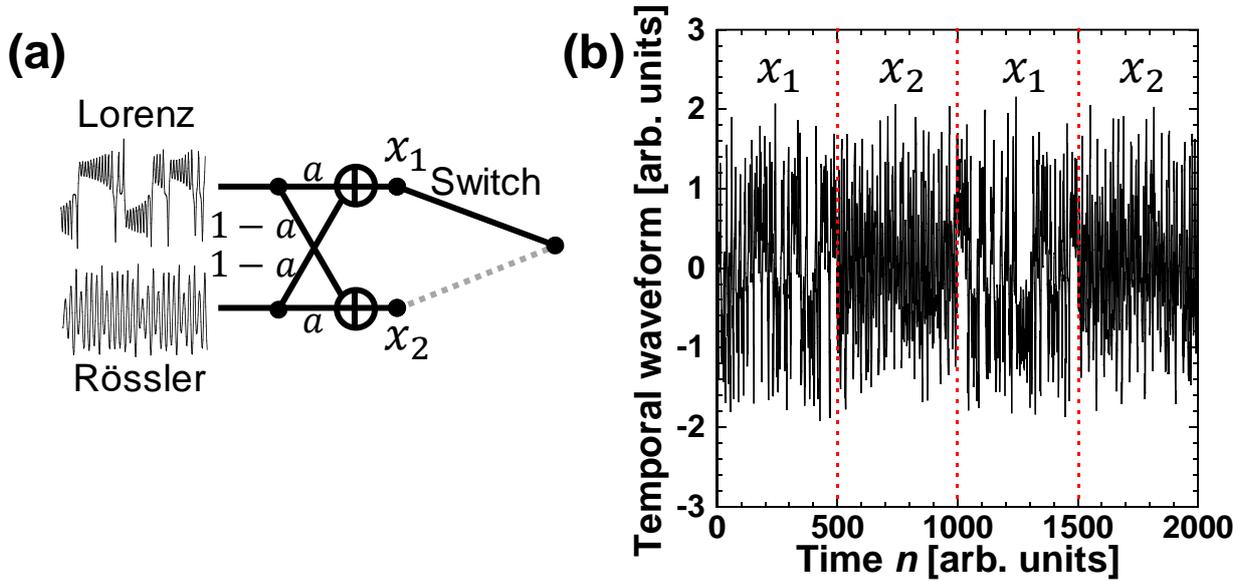

**Figure 6.** (a) Schematic diagram of switching in the mixed time series. The mixing ratio is represented as $a$. (b) Temporal waveform generated by mixing time series from the Lorenz and Rössler models, $x_1 = a\,x_L + (1-a)x_R$ and $x_2 = (1-a)x_L + a\,x_R$ for $a = 0.8$. $x_1$ and $x_2$ are switched every 500 points. $x_1$ is used for $0 \leq n \leq 500$ and $1000 < n \leq 1500$, while $x_2$ is used for $500 < n \leq 1000$ and $1500 < n \leq 2000$.

The temporal evolution of $\Delta e(n)$, $T(n)$, and the selected sequence of the predicted output are summarized in Figs. 7(a), 7(b), and 7(c), respectively. To obtain the mixed time series, $a$ is fixed at 0.8 and the ratio of the Lorenz model is larger than that of the Rössler model. Initially, $\Delta e(n)$ fluctuates around 0, as can be seen in Fig. 7(a), indicating that it would be difficult to identify the correct model. However, the threshold reaches 128 approximately when $n > 35$, as shown in Fig. 7(b), although $\Delta e(n)$ fluctuates around 0. Only $p_2(n)$ is selected after the threshold reaches 128. In other words, correct model selection is achieved in the mixed time series since $p_2(n)$ corresponds to the Lorenz model, whose waveform is dominant in the input signal.

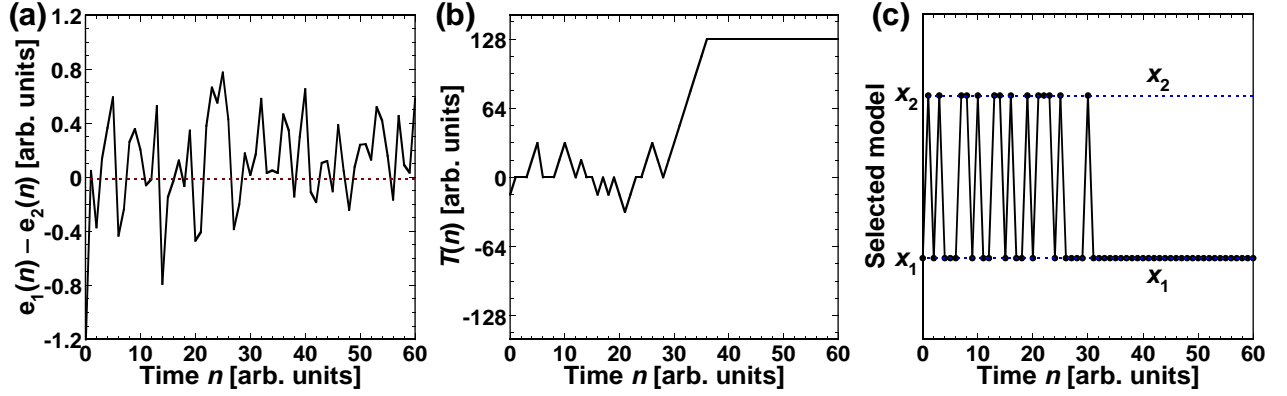

**Figure 7.** (a) Time series of $\Delta e(n) = e_1(n) - e_2(n)$. (b) Temporal dynamics of $T(n)$ for decision making. (c) Models selected by decision making in each time step. $x_1$ is selected for $n > 31$.

$\text{CMSR}(n)$ is calculated to examine the adaptation ability of model selection in the mixed time series. Figure 8(a) shows the temporal evolution of $\text{CMSR}(n)$, and an enlarged view is provided in Fig. 8(b). The target time series is shown in Fig. 6(b), which is obtained by alternating between $x_1$ and $x_2$ every 500 points at $a = 0.8$. The red curve represents $\text{CMSR}(n)$ in the case of the mixed time series. The black curve is the same as that in Fig. 5 and is included for comparison with the mixed time series. The black curve corresponds to the case in which $x_1$ and $x_2$ are switched for $a = 1.0$. $\text{CMSR}(n)$ quickly increases to 1 in both curves after the prediction begins. When the model is switched at $n = 500$, 1,000, and 1,500, $\text{CMSR}(n)$ decreases to 0. However, $\text{CMSR}(n)$ quickly increases to 1 after the switch. Therefore, the correct model is selected adaptively with environmental changes, which means successful model selection. For the mixed time series case, the difference between $e_1(n)$ and $e_2(n)$ fluctuates around 0, as shown in Fig. 7(b). The fluctuation of $\Delta e(n)$ around 0 results in a slower increase of $\text{CMSR}(n)$. However, $\text{CMSR}(n)$ becomes 1, and the correct model is selected successfully.

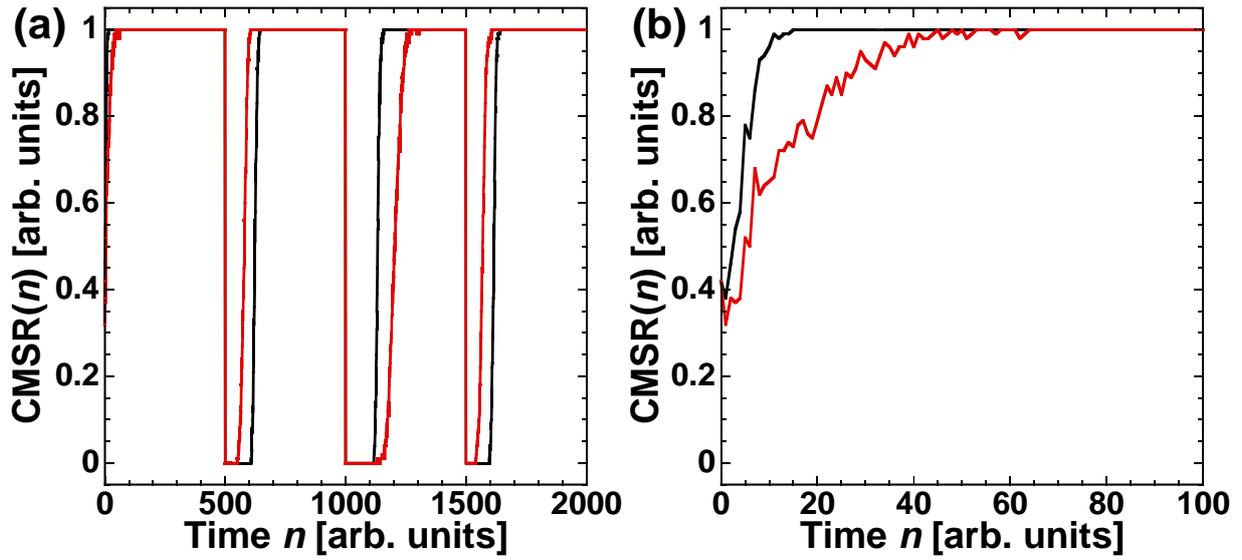

**Figure 8.** (a) Correct model selection rate (CMSR) in the time series prediction task. (b) Enlarged view of (a). The red curve represents the case in which the input signal is a mixed time series consisting of $x_1$ and $x_2$, as shown in Fig. 6(b). For comparison, the black curve represents the case in which the input signal is switched between the Lorenz and Rössler models every 500 steps, as shown in Fig. 2.

The possibility of model selection in the mixed time series is investigated while changing $a$. Figure 9 shows $\text{CMSR}(n)$ at $n = 300$ as a function of $a$. In total, 1,000 trials are conducted in the calculation of $\text{CMSR}(n)$. The target in the model selection is the Rössler model for $a < 0.5$ and the Lorenz model for $a \geq 0.5$. When $a < 0.50$, $\text{CMSR}(n)$ becomes 1 and the target model (Rössler model) is successfully selected. On the other hand, $\text{CMSR}(n)$ does not always reach 1 when $a \geq 0.50$, where the target is the Lorenz model. For a large value of $a$ $(= 0.70)$, $\text{CMSR}(n)$ reaches 1. However, $\text{CMSR}(n) < 1$ when $a < 0.70$. In particular, $\text{CMSR}(n)$ is close to 0.03 when $a = 0.55$, where the Rössler model is selected in most trials. The reason that the Lorenz model is not selected when $a = 0.55$ is that $e_2(n)$ for the Lorenz model is larger than $e_1(n)$ for the Rössler model. The output weights for the Lorenz model are trained from the time series whose attractor has a butterfly structure, and the Lorenz model shows unique oscillations in the upside and downside directions in the time series. The prediction accuracy for the output weights of the Lorenz model decreases if the butterfly structure does not appear in the mixed time series. The butterfly structure can be identified for $a = 0.8$ in Fig. 6. However, this structure cannot be clearly observed for $a < 0.7$ in the mixed time series. Then, the prediction accuracy decreases and the Rössler model is selected when $a = 0.55$.

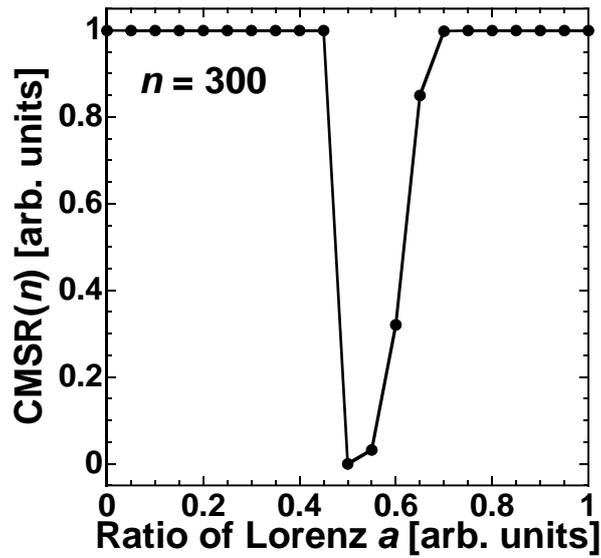

**Figure 9.** Correct model selection rate (CMSR) when $n = 300$ as a function of $a$. The correct selection is the Rössler model for $a < 0.5$ and the Lorenz model for $a \geq 0.5$.

### Adaptive model selection between Rössler models with different parameter values

In the previous two cases, adaptive model selection between two different models (the Rössler and Lorenz models) is investigated. In this subsection, Rössler models with different parameter values are considered, where the parameter value change corresponds to model switching, as shown in Fig. 10(a). This situation of parameter switching is expected to be more difficult than the case of switching between models. Figure 10(b) shows the prediction target. The parameter that is changed in the Rössler model is represented as $b$ (see the *Methods* section) and is switched to $b_1 = 0.2$ and $b_2 = 0.6$, where the Rössler model shows chaotic dynamics in both cases. The first 500 points in the target time series are generated using $b_2 = 0.6$. The switching interval is 500 points, and the total number of points in the time series is 2,000. The weights $w_1$ and $w_2$ are trained for $b_1 = 0.2$ and $b_2 = 0.6$, respectively.

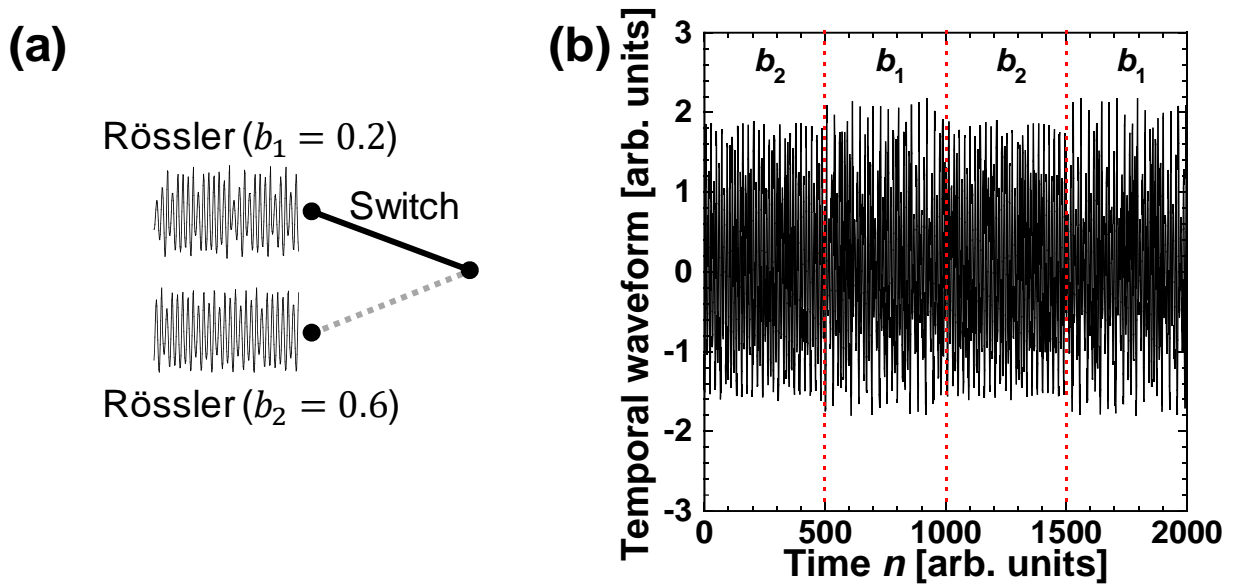

**Figure 10.** (a) Schematic diagram of switching in the case of the Rössler model with different values of $b$. (b) Temporal waveform generated from the Rössler model with $b = b_1 = 0.2$ and $b = b_2 = 0.6$. In the first 500 points of the time series, $b_2$ is used, and switching between $b_2$ and $b_1$ is performed every 500 points.

$\text{CMSR}(n)$ is investigated in the selection of the Rössler models with the two parameter values. Figure 11 shows the temporal evolution of $\text{CMSR}(n)$. The target time series is depicted in Fig. 10(b), where the value of $b$ is switched at $n = 500, 1,000$, and $1,500$. $\text{CMSR}(n)$ quickly increases to 1 after the parameter value is switched. Therefore, adaptive model selection is effectively performed with parameter switching.

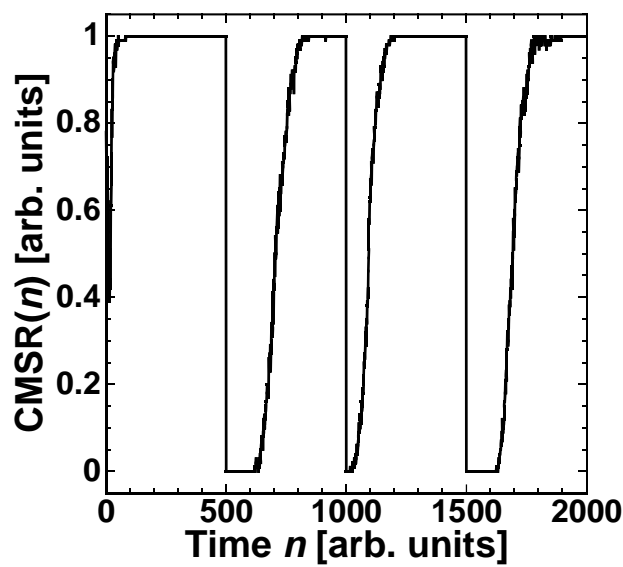

**Figure 11.** Correct model selection rate (CMSR) in the time series prediction task. The target model is the Rössler model with parameter values $b_1$ and $b_2$, and the time series is shown in Fig. 10(b). $b$ is changed every 500 points.

The dependence of model selection on the value of $b$ is investigated using different values of $b_1$, with $b_2$ fixed at 0.6. In this case, the target model is fixed to the Rössler model with $b = b_1$. Figure 12(a) shows $\mathrm{CMSR}(n)$ as a function of $b_1$ at time steps $n = 100$ (black curve) and $n = 300$ (red curve). We focus on how the difference between $b_1$ and $b_2$ is related to the speed of adaptation in model selection. In Fig. 12(a), $\mathrm{CMSR}(n)$ is small near $b_1 = b_2 = 0.6$. $\mathrm{CMSR}(n)$ increases and approaches 1 as the difference between $b_1$ and $b_2$ increases. Therefore, if the two parameter values are apart from each other, the correct model can be selected. In addition, the adaptation speed increases as the difference between $b_1$ and $b_2$ increases.

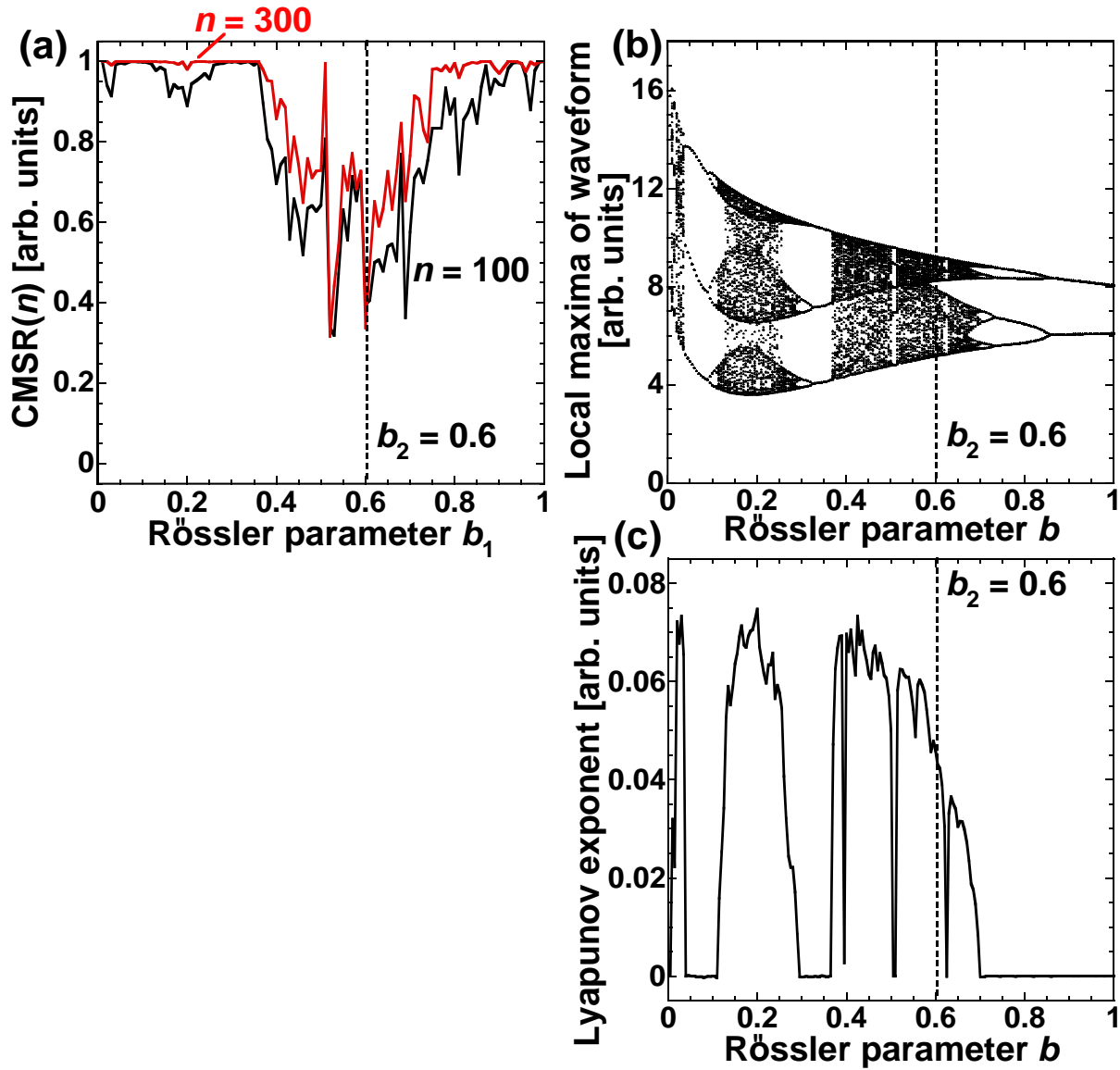

**Figure 12.** (a) Correct model selection rate (CMSR) at $n = 100$ (black curve) and $n = 300$ (red curve) as a function of $b_1$ for the Rössler model. $b_2$ is fixed at 0.6. (b) Bifurcation diagram of the Rössler model as a function of $b$. Local maxima in a time series of $x_R$ are plotted in the bifurcation diagram. (c) The maximum Lyapunov exponent is plotted as a function of $b$.

The temporal dynamics of the Rössler model is also related to the adaptation speed. The bifurcation diagram and maximum Lyapunov exponent of the Rössler model are investigated to determine the dynamics of the Rössler model, as shown in Figs. 12(b) and 12(c), respectively, where $b$ is changed. The bifurcation diagram is generated from the local maxima of the time series of the Rössler model. The Lyapunov exponent quantifies the unpredictability of a dynamic system, and a positive value of the maximum Lyapunov exponent indicates chaotic dynamics. Here, the maximum Lyapunov exponent is positive in three regions: $0 < b < 0.04$, $0.12 < b < 0.26$, and $0.36 < b < 0.70$, except when $b = 0.51$ and $0.625$, where periodic windows are observed. In Fig. 12(a), $\text{CMSR}(n)$ at $n = 100$ does not reach 1 in certain regions of $b_1$ (e.g., $b_1 = 0.2$). However, $\text{CMSR}(n)$ becomes more than 0.99 at $n = 100$ when $0.04 \leq b_1 \leq 0.12$ and $0.26 \leq b_1 \leq 0.36$, where the dynamics is periodic. Although $\text{CMSR}(n)$ at $n = 100$ does not reach 1 when $b_1 \geq 0.70$, it approaches 1 when $n = 300$. Therefore, the adaptation speed is slow if the temporal dynamics at $b = b_1$ and $b = b_2$ are chaotic, and it is fast if the dynamics of the two target models are different (e.g., chaotic and periodic oscillations).

## Conclusions

We proposed an adaptive model selection scheme using reinforcement learning for applications in photonic reservoir computing. Two types of time series were generated using the Rössler and Lorenz models and were exchanged over time to emulate dynamic environmental changes of the incoming signals. We prepared two types of output weights for the Rössler and Lorenz models prior to execution of the prediction task and identified one of the two models for accurate time series prediction using photonic reservoir computing. We succeeded in identifying the correct model adaptively using the prediction errors as rewards in reinforcement learning. The adaptive model selection was also achieved in the case of a mixed time series obtained from the Lorenz and Rössler models with different ratios. We also investigated the adaptive selection of Rössler models with different parameter values. The model selection became easier as the difference between the two parameter values increased. Although two models in reservoir computing were considered in the present study, scalable architecture should be possible; indeed, our former work[29] demonstrated a solution for bandit problems with up to 64 arms using chaotic time series. We consider that constructing a single universal reservoir computing model that can deal with any possible input is most likely impossible; hence, dynamic and autonomous model selection will be a promising means of expanding the computing abilities of photonic artificial intelligence.

## Methods
### Photonic reservoir computing scheme

In reservoir computing, nonlinear mapping of the input information to be processed into a higher-dimensional phase space is required for successful computation[4,5]. A recurrent neural network with a large number of nodes provides the nonlinear mapping in conventional reservoir computing. Instead of using a recurrent neural network, a semiconductor laser and delayed feedback loop can be utilized for photonic reservoir computing[12,30,31]. The reservoir in Fig. 1 consists of a semiconductor laser with delayed optical feedback. A network is virtually emulated by a semiconductor laser and delayed feedback loop. In this scheme, nodes in a network are virtually implemented by temporally dividing the laser output into short time intervals $\theta$, which are called node intervals. Virtual nodes are defined by dividing the delay time of a feedback loop $\tau$ into $\theta$. The number of nodes is given by $N = \tau/\theta$. Therefore, a small value of $\theta$ increases $N$. However, too small of a value of $\theta$ decreases the processing performance of reservoir computing[5]. We used $\theta = 0.1$ ns in this study, and $\tau$ of the reservoir was fixed at $20.0$ ns. Hence, $N = \tau/\theta = 200$.

The input information to be processed is injected into the reservoir after preprocessing the input. We consider discrete time input data $s_n$ ($n = 1, 2, ...$ is the discrete time), which are injected into the reservoir for the duration of $\tau$ to feed the input data to all of the virtual nodes. Before the input data are injected, a mask signal $m(t)$ is multiplied with $s_n$. The mask acts as input weights for virtual nodes and generates transient dynamics in the reservoir. To implement the same input weights for all of the input data, the period of the mask is equal to $\tau$. The mask used in this study was a piece-wise step function with step interval $\theta$. The value of the mask was randomly chosen from the set $\{-1, -0.3, 0.3, 1\}$ (four-level digital mask). The multiplication of the input signal times the mask can be expressed as follows:

$$s(t) = \gamma m(t) s_n \quad ((n-1)T \leq t < nT), \tag{3}$$

where $\gamma$ is the coefficient that scales the amplitude of $s(t)$.

A weighted linear combination of virtual node states is calculated in the output layer, and the calculation result is the output of the RC. The RC output $y(n)$ for the $n$-th input datum is given by the following equation:

$$y(n) = \sum_{j=1}^{N} w_j x_j(n), \tag{4}$$

where $x_j$ is the node state and $w_j$ is the output weight for the $j$-th node state. $x_j$ is extracted from the temporal output of the reservoir, and $w_j$ is trained by minimizing the mean-square error between the target function $\bar{y}(n)$ and RC output $y(n)$ as follows:

$$\frac{1}{N_{tr}} \sum_{n=1}^{N_{tr}} \left(y(n) - \bar{y}(n)\right)^2 \to min, \tag{5}$$

where $N_{tr}$ is the number of input data for training.

**Numerical model for photonic reservoir**

The reservoir is an external cavity semiconductor laser with feedback phase modulation. The temporal dynamics of the laser is described by the Lang-Kobayashi equations[32]:

$$\frac{dE(t)}{dt} = \frac{1 + i\alpha}{2} \left\{ \frac{G_N(N(t) - N_0)}{1 + \epsilon |E(t)|^2} - \frac{1}{\tau_p} \right\} E(t) + \kappa E(t - \tau) \exp[i\{s(t) - \omega\tau\}] + \xi(t) \tag{6}$$

$$\frac{dN(t)}{dt} = J - \frac{N(t)}{\tau_s} - \frac{G_N(N(t) - N_0)}{1 + \epsilon |E(t)|^2} |E(t)|^2, \tag{7}$$

where $E$ is the slowly varying complex electric field amplitude and $N$ is the carrier density. $G_N$ is the gain coefficient; $N_0$ is the carrier density at transparency; $\alpha$ is the linewidth enhancement factor; $\tau_p$ and $\tau_s$ are the photon and carrier lifetimes, respectively; and $J$ is the injection current of the laser. $J$ is given by the product of the lasing threshold current $J_{th}$ and $j$, where $j$ is the normalized injection current. $\omega$ is the angular optical frequency of the laser and is given by $\omega = 2\pi/\lambda$, where $\lambda = 1547$ nm is the optical wavelength of the laser. These parameter values are shown in Table 1.

The second term on the right-hand side of Eq. (6) represents the optical feedback. $\kappa$ is the feedback strength and is given by $\kappa = r_3(1 - r_2)/(r_2 \tau_{in})$, where $r_2$ is the reflectivity of the laser facet and $\tau_{in}$ is the round-trip time in the internal cavity of the laser. $\tau$ is the feedback delay time and is related to the number of virtual nodes $N$ ($= \tau/\theta$). $s(t) - \omega\tau$ represents the phase shift of the feedback light due to phase modulation and feedback delay. $s(t)$ is the input signal for reservoir computing, and the input signal is injected into the reservoir via feedback phase modulation[33]. The last term $\xi(t)$ on the right-hand side of Eq. (6) represents the effect of spontaneous emission noise. $\xi(t)$ is the normalized white Gaussian noise with the properties $\langle \xi(t) \rangle = 0$ and $\langle \xi(t_0)\xi(t) \rangle = \delta(t - t_0)$, where $\langle \cdot \rangle$ denotes the ensemble average and $\delta$ is the Dirac delta function.

## Chaotic dynamical models for generating prediction targets

The prediction targets were generated by the Rössler and Lorenz models, which are well-known models that can generate deterministic chaos. The temporal dynamics of the Rössler and Lorenz models are represented in the following equations. For the Rössler model,

$$\frac{dx_R}{dt} = -y_R - z_R \tag{5}$$

$$\frac{dy_R}{dt} = x_R + 0.2 y_R \tag{6}$$

$$\frac{dz_R}{dt} = b + x_R z_R - 5.7 z_R, \tag{7}$$

For the Lorenz model,

$$\frac{dx_L}{dt} = 10(y_L - x_L) \tag{8}$$

$$\frac{dy_L}{dt} = -x_L z_L + 28 x_L - y_L \tag{9}$$

$$\frac{dz_L}{dt} = x_L y_L - \frac{8}{3} z_L, \tag{10}$$

In the Rössler model, parameter $b$ was set to $0.2$ unless otherwise specified. Variables $x_R$ and $x_L$ were used for the prediction test. The time series of $x_R$ and $x_L$ were normalized with their variances so as not to be identifiable based on knowledge of the amplitudes of the time series.

**Table 1.** Parameter values used in numerical simulations

| Symbol | Parameter | Value |
| --- | --- | --- |
| $G_N$ | Gain coefficient | $8.40 \times 10^{-13}$ m$^3$s$^{-1}$ |
| $N_0$ | Carrier density at transparency | $1.40 \times 10^{24}$ m$^{-3}$ |
| $\varepsilon$ | Gain saturation coefficient | $2.000 \times 10^{-23}$ |
| $\tau_p$ | Photon lifetime | $1.927 \times 10^{-12}$ s |
| $\tau_s$ | Carrier lifetime | $2.04 \times 10^{-9}$ s |
| $\tau_{in}$ | Round-trip time in the internal cavity | $8.0 \times 10^{-12}$ s |
| $r_2$ | Reflectivity of the laser facet | 0.556 |
| $\alpha$ | Linewidth enhancement factor | 3.0 |
| $\lambda$ | Optical wavelength of the laser | $1.537 \times 10^{-6}$ m |
| $c$ | Speed of light | $2.998 \times 10^8$ ms$^{-1}$ |
| $r_3$ | Reflectivity of the external mirror | 0.02 |
| $j$ | Normalized injection current of the laser | 2.00 |
| $\tau$ | Feedback delay time | $20.0 \times 10^{-9}$ s |

**Acknowledgments**

This work was supported in part by JSPS KAKENHI JP19H00868 and JST CREST JPMJCR17N2.

**Author contributions**

All authors have contributed to development and/or implementation of the concept.

K. K. performed the numerical simulations and analyzed the data. K. K., M. N, and A. U. contributed to the discussion of the results. K. K., M. N, and A. U. contributed to the writing of the manuscript.

**Competing Interests**

The authors declare that they have no competing interests.